\documentclass[journals]{IEEEtran}

\usepackage{xspace,amsmath,amssymb,amsfonts,epsfig,syntonly}
\usepackage{cite,bm,color,url,textcomp,empheq,boxedminipage}
\usepackage{graphicx}
\usepackage{algorithmicx,algorithm,algpseudocode}
\usepackage{epstopdf}
\usepackage{empheq}
\usepackage{graphicx,graphics}  
\usepackage{multirow,multicol}
\usepackage{psfrag}    
\usepackage{stfloats}
\usepackage{url}

\newtheorem{lemma}{\underline{Lemma}}[section]

\newtheorem{proposition}{{\underline{Proposition}}}[section]

\newtheorem{remark}{\underline{Remark}}[section]

\long\def\symbolfootnote[#1]#2{\begingroup
\def\thefootnote{\fnsymbol{footnote}}
\footnote[#1]{#2}\endgroup}

\psfull

\allowdisplaybreaks[4]

\begin{document}
\title{Computation Rate Maximization for Multiuser Mobile Edge Computing Systems With \\ Dynamic Energy Arrivals }
\author{Zhifei Lin, Feng Wang, and Licheng Liu\\
School of Information Engineering, Guangdong University of Technology, Guangzhou 510006, China\\
Email: linfei0922@outlook.com, fengwang13@gdut.edu.cn, celcliu@gdut.edu.cn
\vspace{-1.1cm}
}
\maketitle
\begin{abstract}
 This paper considers an energy harvesting (EH) based multiuser mobile edge computing (MEC) system, where each user utilizes the harvested energy from renewable energy sources to execute its computation tasks via computation offloading and local computing. Towards maximizing the system's weighted computation rate (i.e., the number of weighted users' computing bits within a finite time horizon) subject to the users' energy causality constraints due to dynamic energy arrivals, the decision for joint computation offloading and local computing over time is optimized {\em over time}. Assuming that the profile of channel state information and dynamic task arrivals at the users is known in advance, the weighted computation rate maximization problem becomes a convex optimization problem. Building on the Lagrange duality method, the well-structured optimal solution is analytically obtained. Both the users' local computing and offloading rates are shown to have a monotonically increasing structure. Numerical results show that the proposed design scheme can achieve a significant performance gain over the alternative benchmark schemes.
\end{abstract}


\IEEEpeerreviewmaketitle
\symbolfootnote[0]{$^*$F. Wang is the corresponding author.}

\vspace{-0.5cm}
\section{Introduction}
 Driven by the emerging low-latency internet-of-things (IoT) applications building on a large scale of low-power wireless devices, energy harvesting (EH) based mobile edge computing (MEC) systems have recently attracted extensive interests from industry and academia\cite{Mach17,Mao17,Feng19,Sheng16,Chen16,D2D}. In such systems, low-power wireless devices can harness energy from renewable energy sources (such as solar, wind, and radio-frequency signals) and convert it to electrical energy to prolong their lifetime in communication and computation. Powered by the harvested energy, these devices can complete the execution of their computation tasks. Particularly, the computation tasks of each device can be first partitioned into two parts and then be executed by the device's computation offloading and local computing, respectively. For task execution via computation offloading, the devices need to send tasks to the access points (AP) for MEC computing in the uplink channel, and then download the computation results in the downlink channel. By exploiting benefits of both EH and MEC techniques, the EH-based MEC technique is expected to achieve the vision of IoT ubiquitous computing.

 Different from the MEC systems with fixed battery-powered devices, the users' harvested energy from random energy sources is in random and uncertain amounts. Meanwhile, wireless communication channels for computation offloading may be time-varying due to channel fading and mobility. Therefore, such EH-based MEC systems call for new radio/computation resource allocation schemes for computation offloading designs by considering dynamic energy arrivals and channel variation over time. There have many research works on EH-based MEC system designs in the literature \cite{Feng18,Letaief17,Xiao19,Zhang19,Zhang20,Feng20,You16,Bi18,Xu17,Zhou18,TCOM20}. In \cite{Letaief17,Xu17}, EH-based MEC offloading designs for enhancing the long-term system performance were studied based on Lyapunov optimization techniques. In \cite{Feng18,You16,Zhou18,Bi18,Feng20,TCOM20}, binary and partial computation offloading scenarios with wireless power transfer were considered for system energy minimization, and the optimal joint computation-radio resource allocation strategies were derived using a convex optimization framework. Based on the predicated amounts of renewable energy within a given time duration, the works in \cite{Xiao19,Zhang20,Zhang19} investigated reinforcement learning based offloading and energy management schemes for IoT devices with EH.

 Based on the above discussions, there still lacks EH-based MEC designs for maximizing the system computation performance with dynamic energy arrivals over time. In this paper, by focusing on a finite time horizon with multiple equal-length slots, we study a multiuser EH-based MEC system design to maximize the users' weighted computation rate. Suppose that each user harvests energy in a random amount from renewable sources at each slot. Utilizing the harvested energy, the users need to complete the computation of their tasks (via local computing and computing offloading) within the time horizon in a large amount as far as possible. The contribution of this paper is summarized as follows. We first develop a design framework to maximize the users' weighted computing rate (equivalently, the total number of users' computing bits within the finite time horizon) subject to the energy causality constraints over time. The users' local computing and task offloading decisions over different slots are jointly optimized. Then, we pursue an {\em offline} optimization by assuming perfect knowledge of the users' channel state information and energy arrivals is known in advance for revealing engineering insights. In the offline case, the weighted computation rate maximization problem under consideration is a convex optimization problem. Building on the Lagrange duality method, we analytically attain the optimal offline solution, as well as developing a gradient algorithm. At the optimality, each user's local computing rate and offloading rate both have a monotonically increasing structure. Numerical results are also provided to illustrate the benefit of the proposed design scheme over the existing baseline schemes.

 The remainder of the paper is organized as follows. The system model and problem formulation for weighted computation rate maximization are presented in Section~II. Section~III analytically obtains the optimal offline solution. Numerical results are provided to evaluate the proposed scheme in Section~IV, followed by the conclusion in Section~V.

\vspace{-0.2cm}
\section{System Model and Problem Formulation}

\subsection{System Model}
\begin{figure}
  \centering
  \includegraphics[width=2.8in]{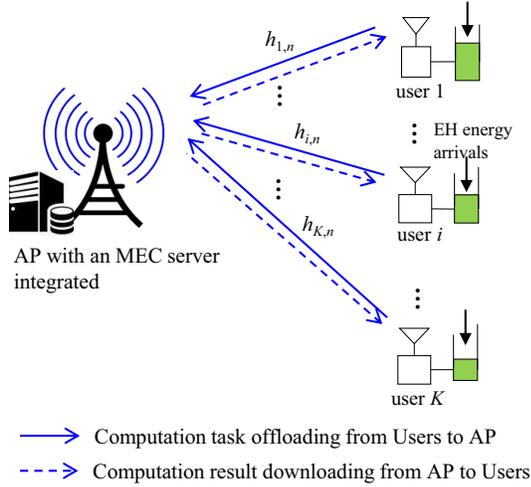}
    \vspace{-0.2cm}
\caption{System model with dynamic energy arrivals at multiple users.}\label{Fig.system}
  \vspace{-0.2cm}
\end{figure}

As shown in Fig.~\ref{Fig.system}, we consider a multiuser EH-based MEC system, where an MEC server is deployed within the AP and each user harvested energy from renewable sources in random amount. We consider a time horizon of length $T>0$, which consists of $N$ equal-length time slots. Powered by the harvested energy, user $k\in{\cal K}\triangleq\{1,...,K\}$ can execute its partitionable computation tasks via local computing and performing task offloading to the AP for MEC execution~\cite{Feng18}.

First, we consider the users' energy consumption in performing local computing by themselves. Denote by $\ell^{\rm loc}_{k,n}$ the number of task input-bits to be executed by user $k$'s local computing within slot $n\in{\cal N}\triangleq\{1,...,N\}$. Based on the dynamic voltage and frequency scaling (DVFS) technique for CPU\cite{Mach17,Mao17,Feng19}, user $k$'s CPU frequency to execute each CPU cycle can then be adjusted as ${C_k \ell^{\rm loc}_{k,n}}/{\tau}$ for the execution of a number of $\ell^{\rm loc}_{k,n}$ task input-bits at slot $n$, where $\tau=T/N$ and $C_k$ denotes the CPU cycle number to compute one task input-bit at user $k$'s CPU. At slot $n\in{\cal N}$, the amount of energy consumed by user $k$'s local computing is then written as
\begin{align}
E_{k,n}^{\rm loc}=C_k \ell^{\rm loc}_{k,n} \gamma_k \Big(\frac{C_k \ell^{\rm loc}_{k,n}}{\tau} \Big)^2 =\frac{\gamma_k C_k^3(\ell^{\rm loc}_{k,n})^3}{\tau^2},~\forall k\in{\cal K},
\end{align}
where $\gamma_k$ denotes a constant effective capacitance coefficient of user $k$'s CPU\cite{Mach17}.

Then, we consider the users' energy consumption in performing task offloading to the AP. Let $\ell^{\rm off}_{k,n}$ denote the number of task input-bits offloaded from user $k\in{\cal K}$ to the AP at slot $n\in{\cal N}$. Denote by $h_{k,n}>0$ the offloading channel power gain from user $k$ to the AP at slot $n$. It is assumed that $\{h_{k,n}\}$ are globally available in this multiuser MEC system e.g., via a pilot-based channel estimation methods. At slot $n$, the achievable transmission rate (in bits/second) for user $k$ to offload tasks to the AP is given by $r_{k,n}=B\log_{2}(1+\frac{p_{k,n} h_{k,n} }{\sigma_0^2})$, $\forall n\in{\cal N}$, where $p_{k,n}$ represents user $k$'s transmit power at slot $n$, $B$ is the bandwidth, and $\sigma_0^2$ is the zero-mean additive white Gaussian noise (AWGN) variance at the AP receiver. As such, it yields that $\ell^{\rm off}_{k,n}=r_{k,n}\tau$, $\forall k\in{\cal K},n\in{\cal N}$. At slot $n\in \mathcal N$, the amount of energy consumed by user $k$ in performing task offloading to the AP is given by
\begin{align}
E_{k,n}^{\rm off} = \tau p_{k,n} = \frac{\sigma_0^2\tau}{h_{k,n}}\left(2^{\frac{\ell^{\rm off}_{k,n}}{\tau B}}-1\right),~\forall k\in{\cal K}.
\end{align}

Next, we model the energy causality constraints for the users due to energy harvesting based on renewable sources. Specifically, the cumulative amount of energy consumed by user $k$ in performing local computing and offloading at slot $n$ is no more than that of its harvested energy until this slot $n$. Let $E_{k,n}$ denote the amount of energy harvested by user $k$ at the beginning of slot $n\in \mathcal{N}\setminus\{N\}$. It is assumed that the rechargeable battery's capacity of each user is sufficiently large, such that all the harvested energy can be safely stored. Consider the energy harvested by each user at a current slot can be immediately utilized in the next slots. As a result, the energy causality constraints at user $k\in{\cal K}$ are expressed as
\begin{align}\label{eq.energy-causality}
\sum_{j=1}^{n}E_{k,j}^{\rm loc} + \sum_{j=1}^n E_{k,j}^{\rm off}\leq E_{k,0}+\sum_{j=1}^{n-1}E_{k,j},~ \forall n\in{\cal N},
\end{align}
where $E_{k,0}>0$ denotes user $k$'s initial amount of energy stored in its rechargeable battery before energy harvesting.

\subsection{Problem Formulation}
In this paper, our objective is to maximize the $K$ users' weighted computation rate under the energy causality constraints \eqref{eq.energy-causality} over the $N$ slots. This corresponds to maximizing the weighted number of task input-bits executed by all the users within the finite horizon. We jointly optimize the number of task input-bits for local computing and task offloading per slot at the users. Let $\omega_k>0$ denote the computation weight of user $k$, which characterizes user $k$'s priority or preference. Therefore, the weighted computation rate maximization problem under consideration is formulated as
\begin{subequations}\label{eq.prob1}
\begin{align}
({\rm P}1):&\max_{\{ \ell^{\rm loc}_{k,n}\geq 0,~\ell^{\rm off}_{k,n}\geq 0\}}~~\sum_{k=1}^K\sum_{n=1}^N
\omega_k(\ell^{\rm loc}_{k,n}+\ell^{\rm off}_{k,n})\\
&\mathrm{s.t.}~~\sum_{j=1}^{n} E_{k,j}^{\rm loc} + \sum_{j=1}^{n} E_{k,j}^{\rm off} \leq E_{k,0}+\sum_{j=1}^{n-1} E_{k,j},\notag \\
&\quad\quad\quad\quad\quad\quad\quad\quad\quad\quad\quad~~\forall k\in {\cal K},n\in{\cal N},
\end{align}
\end{subequations}
where the energy causality constraints in (\ref{eq.prob1}b) specify that the cumulative energy amount at each slot used for both local computing and offloading must be less than or equal to that of energy available (including the harvested and initial energy) for each user. Note that both the profiles $\{h_{k,n},E_{k,n}\}_{n=1}^N$ of channel state information (CSI) and energy state information (ESI) are not necessarily obtained for each user $k\in{\cal K}$. As such, problem (P1) is difficult to solve. To reveal the optimal decisions for task execution over time, we consider the offline case when the CSI/ESI profile $\{h_{k,n},E_{k,n}\}$ is known in advance. In the offline case, (P1) is a convex optimization problem, and can then be efficiently solved by generic convex solvers (e.g., CVX toolbox)\cite{Boyd2004}. Alternatively, we next employ the Lagrangian duality method to analytically attain the optimal solution of problem (P1).

\section{Optimal Offline Solution to Problem (P1)}

In this section, we present the optimal offline solution of (P1) based on the celebrated Lagrange duality method \cite{Boyd2004}.

\vspace{-0.2cm}
\subsection{Lagrange Dual Problem of (P1)}
Let $\mu_{k,n} \geq 0$ denote the non-negative Lagrange multiplier associated with the ($k,n$)-th constraint in (\ref{eq.prob1}b), $\forall k\in{\cal K}$, $n\in{\cal N}$. The {\em partial} Lagrangian of problem (P1) is given by\cite{Boyd2004}
\begin{align}\label{eq.F}
& {\cal F}(\{\ell^{\rm loc}_{k,n},\ell^{\rm off}_{k,n},\mu_{k,n}\}) =\sum_{k=1}^K \sum_{n=1}^N
\omega_{k}(\ell^{\rm loc}_{k,n}+\ell^{\rm off}_{k,n}) \notag \\
 & +\sum_{k=1}^K \sum_{n=1}^N \mu_{k,n} \Bigg(E_{k,0} + \sum_{j=1}^{n-1} E_{k,j} - \sum_{j=1}^n \frac{\gamma_kC_k^3(\ell^{\rm loc}_{k,j})^3}{\tau^2} \notag \\
 &\quad\quad\quad\quad\quad\quad\quad\quad - \sum_{j=1}^n \frac{\sigma_0^2\tau}{h_{k,n}}(2^{\frac{\ell^{\rm off}_{k,j}}{\tau B}}-1 ) \Bigg) \notag \\
&=\sum_{k=1}^K\sum_{n=1}^N\Bigg( \omega_k\ell_{k,n}^{\rm loc} - \Big(\sum_{j=n}^N \mu_{k,j}\Big)\frac{\gamma_i C_k^3(\ell_{k,n}^{\rm loc})^3}{\tau^2} \Bigg)\notag \\
&~~ +\sum_{k=1}^K\sum_{n=1}^N\Bigg( \omega_k \ell_{k,n}^{\rm off} -\Big(\sum_{j=n}^N \mu_{k,j}\Big)\frac{\sigma_0^2\tau}{h_{k,n}}\big(2^{\frac{\ell_{k,n}^{\rm off}}{\tau B}}-1\big) \Bigg) \notag\\
&~~ + \sum_{k=1}^K\sum_{n=1}^{N}\mu_{k,n}E_{k,0} + \sum_{k=1}^K\sum_{n=1}^{N-1} \Big(\sum_{j=n+1}^{N} \mu_{k,j}\Big)E_{k,n}.
\end{align}
Based on \eqref{eq.F}, the Lagrange dual problem associated with (P1) is obtained as
\begin{align}
({\rm D}1):&\min_{\{ \mu_{k,n}\geq 0\}} {\cal G}(\{\mu_{k,n}\}),
\end{align}
where the so-called dual function ${\cal G}(\{\mu_{k,n}\})$ is defined as the maximum value of ${\cal F} (\{\ell^{\rm loc}_{k,n},\ell^{\rm off}_{k,n},\mu_{k,n}\})$ over the primal variables $(\{\ell^{\rm loc}_{k,n},\ell^{\rm off}_{k,n}\})$\cite{Boyd2004}, i.e.,
\begin{align}\label{eq.dual-func}
{\cal G}(\{\mu_{k,n}\}) \triangleq  \max_{\{ \ell^{\rm loc}_{k,n}\geq 0,~\ell^{\rm off}_{k,n}\geq 0\}} {\cal F} (\{\ell^{\rm loc}_{k,n},\ell^{\rm off}_{k,n},\mu_{k,n}\}).
\end{align}

In the following, we solve problem (P1) by first Obtaining ${\cal G}(\{\mu_{k,n}\})$ in (5) under given $\mu_{k,n}\geq 0$, $\forall k\in{\cal K}$, $n\in{\cal N}$, and then solving problem (D1) to find the optimal $\mu_{k,n}$. Denote by $(\{ \ell^{\rm loc}_{k,n})^*,(\ell^{\rm off}_{k,n})^*\})$ the optimal solution of \eqref{eq.dual-func} under given $\mu_{k,n}\geq 0$, and let $(\{(\ell^{\rm loc}_{k,n})^{\rm opt},(\ell^{\rm off}_{k,n})^{\rm opt}\})$ and $\{\mu_{k,n}^{\rm opt}\}$ denote the optimal primal and dual solutions for (P1) and (D1), respectively.

\subsection{Dual Decomposition and Evaluation of ${\cal G}(\{\mu_{k,n}\})$}
First, we evaluate the dual function ${\cal G}(\{\mu_{k,n}\})$ in \eqref{eq.dual-func} under the given $\{\mu_{k,n}\geq 0\}$. Note that dual problem \eqref{eq.dual-func} can be decomposed into $2KN$ univariate subproblems as follows.
\begin{align}\label{eq.prob-loc}
\max_{\ell^{\rm loc}_{k,n}\geq 0}~~\omega_k\ell_{k,n}^{\rm loc} - \Big(\sum_{j=n}^N \mu_{k,j}\Big)\frac{\gamma_k C_k^3(\ell_{k,n}^{\rm loc})^3}{\tau^2}
\end{align}
\begin{align}\label{eq.prob-off}
\max_{\ell^{\rm off}_{k,n}\geq 0}~~\omega_k\ell_{k,n}^{\rm off} -\Big(\sum_{j=n}^N \mu_{k,j}\Big)\frac{\sigma_0^2\tau}{h_{k,n}}\Big(2^{\frac{\ell_{k,n}^{\rm off}}{\tau B}}-1\Big),
\end{align}
where $k\in{\cal K}$ and $n\in{\cal N}$. Before solving problem in \eqref{eq.prob-loc} and \eqref{eq.prob-off}, we establish the following lemma.
\begin{lemma}[Positivity of $\mu_{k,N}$]\label{Lem:Pos}
At the optimality of dual problem (D1), there always exists a positive Lagrange multiplier $\mu_{k,N}$ for $k\in{\cal K}$, i.e., $\mu_{k,N}>0$.
\end{lemma}
\begin{IEEEproof}
Lemma~\ref{Lem:Pos} can be verified by checking the complementary slackness conditions, i.e.,
\begin{align*}
 \mu_{k,N} \Big(E_{k,0}+\sum_{j=1}^{n-1}E_{k,n} - \sum_{j=1}^{n} E_{k,j}^{\rm loc} - \sum_{j=1}^{n} E_{k,j}^{\rm off}\Big)=0, \forall k\in{\cal K}.
\end{align*}
In order to maximize the weighted number of computation task input-bits across the users for problem (P1), by contradiction it shows that the $N$-th energy causality constraint (\ref{eq.prob1}b) for each user $k\in{\cal K}$ must be active at the optimality; i.e., it always holds that $\sum_{j=1}^{N} E_{k,j}^{\rm loc} + \sum_{j=1}^{N} E_{k,j}^{\rm off} = E_{k,0}+\sum_{j=1}^{N-1} E_{k,j}$ for all $k\in {\cal K}$. Therefore, one can always set $\mu_{k,N}>0$ without violating the complementary slackness conditions.
\end{IEEEproof}

Based on Lemma~\ref{Lem:Pos}, it yields that $\sum_{j=n}^N\mu_{k,n}>0$, $\forall k\in{\cal K}$, $n\in{\cal N}$. As a result, all the objective functions in problems \eqref{eq.prob-loc} and \eqref{eq.prob-off} are convex and bounded above. Based on the Karush-Kuhn-Tucker (KKT) conditions\cite{Boyd2004}, we explicitly obtain their optimal $(\ell_{k,n}^{\rm loc})^*$ and $(\ell_{k,n}^{\rm off})^*$, respectively, as stated in the following proposition.

\begin{proposition}[Primal Solution $(\{(\ell^{\rm loc}_{k,n})^*,(\ell^{\rm off}_{k,n})^*\})$ of \eqref{eq.dual-func}]\label{prop1} 
For any given $\{\mu_{k,n}\geq 0\}$, the optimal number of task input-bits $\{(\ell^{\rm loc}_{k,n})^*\}$ for local computing for \eqref{eq.prob-loc} and $\{(\ell^{\rm off}_{k,n})^*\}$ for task offloading for \eqref{eq.prob-off} are
\begin{subequations}\label{eq.opt-loc-off}
\begin{align}
(\ell^{\rm loc}_{k,n})^* &= \sqrt{\frac{\omega_k\tau^2}{3\left(\sum_{j=n}^N\mu_{k,j}\right)\gamma_k C_k^3}} \\
(\ell^{\rm off}_{k,n})^* &= \tau B \log_2\left(\frac{ \omega_k B h_{k,n}} {\left(\sum_{j=n}^N\mu_{k,j}\right)\sigma_0^2 \ln{2}}\right).
\end{align}
\end{subequations}
\end{proposition}

\begin{IEEEproof}
The optimal $\{(\ell^{\rm loc}_{k,n})^*\}$ and $\{(\ell^{\rm off}_{k,n})^*\}$ are obtained by setting the first-order derivative conditions for the objective functions in \eqref{eq.prob-loc} and \eqref{eq.prob-off} to be zero, respectively.
\end{IEEEproof}

\begin{remark}
From \eqref{eq.opt-loc-off}, due to the nonnegativity of $\{\mu_{k,n}\}$, it follows that the term $\sum_{j=n}^N\mu_{k,j}$ decreases with the increasing of slot index $n$, i.e., $\sum_{j=1}^N\mu_{k,j}\geq \sum_{j=2}^N\mu_{k,j}\geq ... \geq \sum_{j=N-1}^N\mu_{k,j}\geq \sum_{j=N}^N\mu_{k,j}$. Therefore, the optimal number of task input-bits by each user's local computing and offloading (under constant channel power gains) are {\em monotonically increasing} over time, respectively; i.e, $(\ell^{\rm loc}_{k,1})^*\leq ...\leq (\ell^{\rm loc}_{k,N})^*$ and $(\ell^{\rm off}_{k,1})^*\leq ...\leq (\ell^{\rm off}_{k,N})^*$, $\forall k\in{\cal K}$. In order for exploiting the available energy to maximize the executed number of task input-bits, each user should execute its computation tasks as evenly as possible in amount within the horizon. Since a large amount of available energy will be accumulated for each user as time goes on, each user should execute an increasing number of computation task input-bits over time. This monotone structure is also reminiscent of the staircase power allocation strategy in EH-based communication systems for throughput maximization\cite{Ozel11,Zhang12}.
\end{remark}

\subsection{Obtaining Optimal $\{\mu_{k,n}^{\rm opt}\}$ to Minimize ${\cal G}(\{\mu_{k,n}\})$}
With $(\{(\ell_{k,n}^{\rm loc})^*,(\ell_{k,n}^{\rm off})^*\})$ obtained, we next solve the dual problem (D1) to minimize ${\cal G}(\{\mu_{k,n}\})$. Note that ${\cal G}(\{\mu_{k,n}\})$ is a convex and differentiable function in general. Therefore, we use an iterative gradient method to find the optimal $\{\mu_{k,n}^{\rm opt}\}$ for problem (D1)\cite{sub-gradient}. Specifically, we update the dual variables $\{\mu_{k,n}\}$ according to \begin{align} \label{eq.dual-update}
\mu^{(q+1)}_{k,n} = \left( \mu^{(q)}_{k,n} - \eta_k^{(q)} g_{k,n}(\mu^{(q)}_{k,n}) \right)^+,
\end{align}
where $\mu^{q}_{k,n}$ denotes the dual variable at the $q$th iteration, $g_{k,n}(\mu^{(q)}_{k,n})\triangleq E_{k,0} + \sum_{j=1}^{n-1} E_{k,j} - \sum_{j=1}^n \frac{\gamma_kC_k^3(\ell^{\rm loc}_{k,j})^3}{\tau^2} - \sum_{j=1}^n \frac{\sigma_0^2\tau}{h_{k,n}}(2^{\frac{\ell^{\rm off}_{k,j}}{\tau B}}-1 ) $ is the gradient of ${\cal G}(\{\mu_{i,n}\})$ with respect to $\mu^{(q)}_{k,n}$, and $\eta^{(q)}_k>0$ is the $q$th iterative step size. When the difference between two consecutively iterated dual function values is smaller than a certain threshold, the iteration procedure of the gradient method will terminate and the current updated dual variables $\{\mu_{k,n}\}$ are chosen as the optimal dual solution $\{\mu^{\rm opt}_{k,n}\}$.

 \subsection{Finding Optimal $(\{(\ell_{k,n}^{\rm loc})^{\rm opt},(\ell_{k,n}^{\rm off})^{\rm opt}\})$ for (P1)}
 With $\{\mu^{\rm opt}_{k,n}\}$ obtained, we proceed to find the optimal solution $(\{(\ell_{k,n}^{\rm loc})^{\rm opt},(\ell_{k,n}^{\rm off})^{\rm opt}\})$ to problem (P1). Replacing $\{\mu_{k,n}\}$ with $\{\mu^{\rm opt}_{k,n}\}$ in Proposition~\ref{prop1}, we obtain the optimal $(\{(\ell_{k,n}^{\rm loc})^{\rm opt},(\ell_{k,n}^{\rm off})^{\rm opt}\})$ for problem (P1).

 In summary, Algorithm 1 is presented for optimally solving the weighted computation rate maximization problem (P1).

\begin{algorithm}
\caption{for Optimally Solving Problem (P1)}
\begin{algorithmic}[1] 
\State {\bf Initialization:~} Given initial dual variable $\mu^{(0)}_{k,n}>0$, $\forall k\in{\cal K}$, $n\in{\cal N}$ and the prescribed accuracy $\epsilon$, and set iteration number $q=0$ and step size $\eta_k^{(0)}=1$.
\State {\bf Repeat:}
\begin{itemize}
\item For each user $k\in{\cal K}$ and each slot $n\in{\cal N}$, obtain $(\ell^{\rm loc}_{k,n})^*$ and $(\ell^{\rm off}_{k,n})^*$) by Proposition~\ref{prop1} with $\mu_{k,n}\geq 0$;
\item Obtain the gradient $g(\mu^{(q)}_{k,n}) \gets E_{k,0} + \sum_{j=1}^{n-1} E_{k,j} - \sum_{j=1}^n \frac{\gamma_kC_k^3(\ell^{\rm loc}_{k,j})^3}{\tau^2} - \sum_{j=1}^n \frac{\sigma_0^2\tau}{h_{k,n}}(2^{\frac{\ell^{\rm off}_{k,j}}{\tau B}}-1 )$ at the point $\mu^{(q)}_{k,n}$, $\forall k,n$;
\item Update the dual variables $\mu^{(q+1)}_{k,n} \gets \Big( \mu^{(q)}_{k,n} - \eta_k^{(q)} g_{k,n}(\mu^{(q)}_{k,n}) \Big)^{+}$ by \eqref{eq.dual-update};
\item Set $q\gets q+1$ and $\eta^{(q)}_k \gets 1/q$;
\end{itemize}

\State {\bf Until} $\left|{\cal G}(\{\mu^{(q)}_{k,n}\}) - {\cal G}(\{\mu^{(q-1)}_{k,n}\})\right|/{\cal G}(\{\mu^{(q)}_{k,n}\}) < \epsilon$.

\State {\bf Set}~$\mu^{(\rm opt)}_{k,n} \gets
\mu^{(q)}_{k,n}$, $\forall k\in{\cal K},n\in{\cal N}$;
\State {\bf Output:} Obtain $(\ell_{k,n}^{\rm loc})^{\rm opt}$ and $(\ell_{k,n}^{\rm off})^{\rm opt}$ by replacing $\mu_{k,n}$ with $\mu^{({\rm opt})}_{k,n}$ in Proposition~\ref{prop1}.
\end{algorithmic}
\end{algorithm}

\section{Numerical Results}
In this section, we numerically gauge the proposed scheme for multiuser EH-based MEC system. We include the following three baseline schemes for performance comparison.

\begin{itemize}
\item{\em Equal-Energy Allocation Scheme:} The amount of energy harvested at each slot is divided into two identical parts for local computing and offloading, respectively; i.e., $E^{\rm loc}_{k,n}=E^{\rm off}_{k,n}=E_{k,n}/2$, $\forall k\in{\cal K}$, $n\in{\cal N}\setminus\{N\}$.

\item {\em Local Computing Only Scheme:} The amount of harvested energy is used for the users' local computing; i.e., it corresponds to (P1) with $\ell^{\rm off}_{k,n}=0$, $\forall k\in{\cal K}$, $n\in{\cal N}$.

\item{\em Full Offloading Scheme:} The amount of harvested energy is used for the users' task offloading towards the AP; i.e., it corresponds to (P1) with $\ell^{\rm loc}_{k,n}=0$, $\forall k\in{\cal K}$, $n\in {\cal N}$.
\end{itemize}

In the simulations, we set the initial amount of energy to be $E_{k,0}=0.3$~Joule, $\forall k\in{\cal K}$. The amount of each user's harvested energy at slot $n$ is set to be distributed with a uniform distribution $E_{k,n}\in {\cal U}[0,E_0]$, where $E_0=1$ Joule and $n\in{\cal N}\setminus\{N\}$. We set the horizon duration as $T=0.2$ sec and the user weight as $\omega_k=1$, $\forall k\in{\cal K}$, unless specified otherwise. For local computing, we set the CPU switch capacitance coefficient and the number of CPU cycles for executing one task input-bit as $\gamma_k=10^{-28}$ and $C_k=500$, respectively. The AP receiver noise power is $\sigma_0^2=10^{-9}$ Watt. For computation offloading, the bandwidth is set as $B=2$ MHz, and the channel power gain is set as $h_{k,n}=\gamma_0d_k^{-3.5}|\bar{h}_{k,n}|^2$, where $\bar{h}_{k,n}\sim{\cal CN}(0,1)$ captures the small-scale fading effect, $\gamma_0=-50$ dBm denotes the reference pathloss at one meter, and $d_k$ denotes the distance between user $k\in{\cal K}$ to the AP.

\begin{figure}
  \centering
 \includegraphics[width=3.2in]{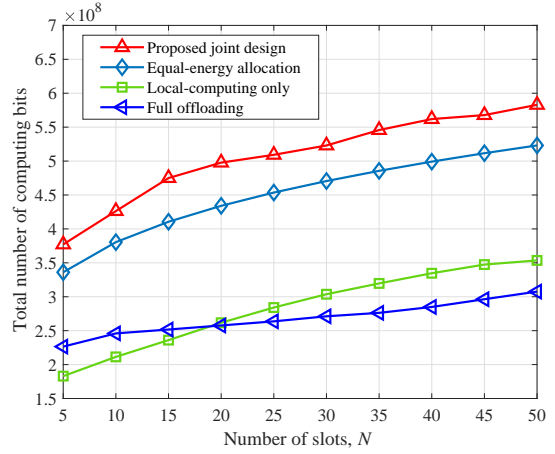}  \vspace{-0.2cm}
  \caption{The total number of users' executed task input-bits versus $N$.} \label{fig.compare}
    \vspace{-0.4cm}
\end{figure}

Fig.~\ref{fig.compare} shows the total number of the $K$ users' computing bits versus the slot number $N$, where $K=50$, the slot duration $\tau=0.02$ sec, and $d_k=20$ meters, $\forall k\in{\cal K}$. As $N$ increases, the total number of task input-bits to be executed by the four schemes increases, and the proposed scheme achieves a significant gain over the benchmark schemes. The benchmark equal-energy-allocation scheme is observed to outperform the other two benchmark schemes. It illustrates the importance of simultaneously exploiting the users' local computing and offloading functionalities to enhance the system computation capability. Also, the local-computing-only scheme outperforms the full-offloading scheme at large $N$ values (e.g., $N>20$), but it is not true when $N$ becomes smaller in this setup.

\begin{figure}
  \centering
  \includegraphics[width=3.2in]{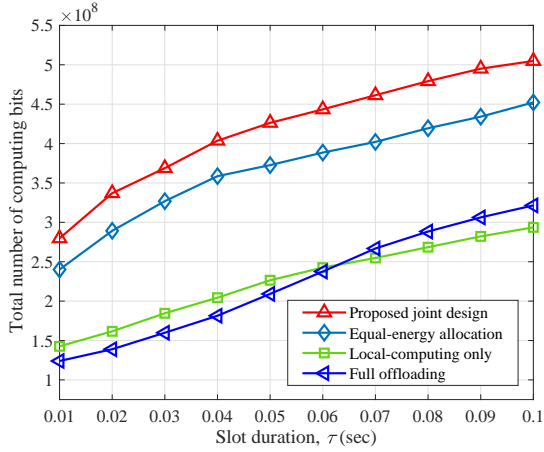}
  \vspace{-0.2cm}
  \caption{The total number of users' executed task input-bits versus $\tau$.} \label{fig.duration}
    \vspace{-0.4cm}
\end{figure}

Fig.~\ref{fig.duration} shows the total number of executed task input-bits versus the slot duration $\tau$, where $K=50$, $N=20$, and $d_k=20$ meters, $\forall k\in{\cal K}$. Again, the proposed scheme achieves a substantial performance gain over the benchmark schemes. The equal-energy-allocation scheme is observed to outperform the other benchmark schemes, which further indicates the benefit of joint local computing and offloading. The full-offloading scheme outperforms the local-computing-only scheme when the slot duration $\tau$ becomes larger (e.g., $>0.06$ sec). This implies the users prefer to offload tasks to the AP at large $\tau$ values to fully utilize the limited energy.

\begin{figure}
  \centering
  \includegraphics[width=3.2in]{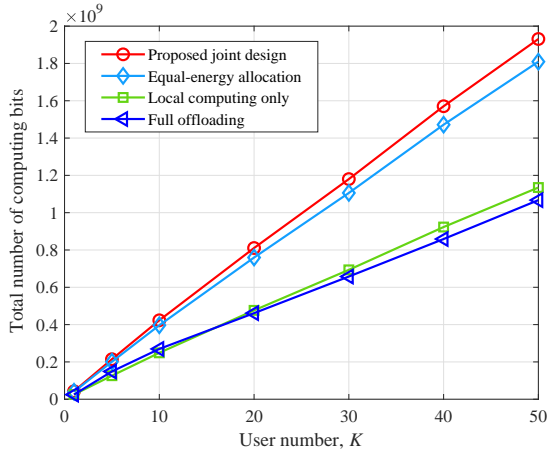}  \vspace{-0.2cm}
\caption{The total number of executed task input-bits versus user number $K$.} \label{fig.tradeoff}
  \vspace{-0.4cm}
\end{figure}

Fig.~\ref{fig.tradeoff} shows the total number of users' executed task input-bits versus the user number $K$, where $N=20$, $\tau=0.02$ sec, and $d_k=20$ meters, $\forall k\in{\cal K}$. It is observed that the number of computing bits of all the schemes increases as $K$ increases. As compared to the local-computing-only and full-offloading schemes, the proposed scheme achieves a significant performance gain, especially when $K$ becomes larger. Again, it indicates the necessity to exploit both local computing and offloading functionalities for maximizing the MEC system performance.

%
\vspace{-0.4cm}
\section{Conclusion}
This paper investigated a multiuser EH-based MEC system design to maximize the users' weighted computation rate over a finite horizon, by taking into account the users' dynamical energy arrivals by energy harvesting. Subject to the energy causality constraints, we jointly optimized the task allocation for the users' local computing and offloading to the AP over time in an offline fashion. Based on the convex optimization methods, the optimal offline solution was efficiently obtained. It revealed that the rates for users' local computing and task offloading have a monotone structure. Numerical results showed the substantial gains of the proposed scheme in maximizing the users' weighted number of executed task input-bits over other benchmark schemes. 



\end{document}